\documentclass[a4paper, 10pt, conference]{IEEEtran}
\IEEEoverridecommandlockouts
\usepackage{cite}
\usepackage{amsmath,amssymb,amsfonts}
\usepackage{algorithmic}
\usepackage{graphicx}
\usepackage{textcomp}
\usepackage{xcolor}
\usepackage{fancyhdr} 
\usepackage{pgfplots}
\usetikzlibrary{patterns}
\pgfplotsset{compat=1.17}
\usepackage{svg}
\usepackage{stfloats}
\usepackage{wrapfig}
\usepackage{xurl}
\usepackage{balance}
\usepackage{subfigure}
\usepackage{tcolorbox}
\usepackage{array}
\def\BibTeX{{\rm B\kern-.05em{\sc i\kern-.025em b}\kern-.08em
    T\kern-.1667em\lower.7ex\hbox{E}\kern-.125emX}}

\title{Characterizing Encrypted Application Traffic through Cellular Radio Interface Protocol\\
}

\makeatletter
\newcommand{\linebreakand}{%
  \end{@IEEEauthorhalign}
  \hfill\mbox{}\par
  \mbox{}\hfill\begin{@IEEEauthorhalign}
}
\makeatother

\author{\IEEEauthorblockN{Md Ruman Islam}
\IEEEauthorblockA{\textit{University of Nebraska Omaha}\\
mdrumanislam@unomaha.edu}
\and
\IEEEauthorblockN{Raja Hasnain Anwar}
\IEEEauthorblockA{\textit{University of Massachusetts Amherst}\\
ranwar@umass.edu} 
\linebreakand
\IEEEauthorblockN{Spyridon Mastorakis}
\IEEEauthorblockA{\textit{University of Notre Dame}\\
mastorakis@nd.edu}
\and
\IEEEauthorblockN{Muhammad Taqi Raza}
\IEEEauthorblockA{\textit{University of Massachusetts Amherst}\\
taqi@umass.edu}
}

\begin{document}
\makeatletter
\patchcmd{\@maketitle}
 {\addvspace{0.5\baselineskip}\egroup}
 {\addvspace{-2\baselineskip}\egroup}
 {}
 {}
\makeatother

\maketitle

\begin{abstract}
Modern applications are end-to-end encrypted to prevent data from being read or secretly modified. 5G technology provides ubiquitous access to these applications without compromising the application-specific performance and latency goals. In this paper, we empirically demonstrate that 5G radio communication becomes the side channel to precisely infer the user's applications in real-time. The key idea lies in observing the 5G physical and MAC layer interactions over time that reveal the application's behavior. The MAC layer receives the data from the application and requests the network to assign the radio resource blocks. The network assigns the radio resources as per application requirements, such as priority, Quality of Service (QoS) needs, amount of data to be transmitted, and buffer size. The adversary can passively observe the radio resources to fingerprint the applications. We empirically demonstrate this attack by considering four different categories of applications: online shopping, voice/video conferencing, video streaming, and Over-The-Top (OTT) media platforms. Finally, we have also demonstrated that an attacker can differentiate various types of applications in real-time within each category.
\end{abstract}


\section{Introduction}
\label{sec:introduction}

\fancyhf{}
\fancyfoot[C]{%
 \parbox{\textwidth}{%
  \textcolor{red}{This paper has been accepted for publication by the $21^\text{st}$ IEEE International Conference on Mobile Ad-Hoc and Smart Systems (MASS 2024). \textcopyright\ 2024 IEEE. Personal use of this material is permitted. Permission from IEEE must be obtained for all other uses, in any current or future media, including reprinting/republishing this material for advertising or promotional purposes, creating new collective works, for resale or redistribution to servers or lists, or reuse of any copyrighted component of this work in other works.}
 }
}
\thispagestyle{fancy}

Despite the widespread use of encryption in network communications today, there are still threats related to the security and privacy of users. As applications on mobile devices connect to the Internet, their data communications over cellular networks create unique fingerprints. These fingerprints can be exploited to infer user activities and applications, potentially putting users' sensitive information in the hands of adversaries and elevating security and privacy risks even further~\cite{r22, r3, r2}. In this paper, we perform an empirical study on characterizing users' activities in the wild despite these applications being end-to-end encrypted.

We discover that the interactions between the 5G device MAC layer and the physical layer reveal what sort of applications the device is running. When the device application layer sends some data, it pushes it to the 5G PDCP layer~\cite{alba2019traffic, schaich2018antenna}. The PDCP protocol transfers the data to the MAC layer, which stores the data into an application-specific buffer for its transmission over the physical layer. The MAC scheduler communicates with the 5G base station (via the physical layer) and requests the allocation of radio resources, such as Radio Resource Blocks (RRBs), according to application requirements, such as Quality of Service (QoS), priority, and buffer sizes~\cite{mamane2022scheduling}. 
For example, if the device has voice call data packets to send, the 5G base station uses persistent scheduling to allocate a fixed number of RBs to the device at every physical layer subframe; however, in the case of HTTP packets, the base station might use dynamic scheduling approach to allocate resources to the device according to data packets in the buffer. 
The adversary observes this interaction and fingerprints the application behavior in real-time. 

\begin{figure}
    \centering
    \subfigure[Shopping Websites]{
    \includegraphics[width=0.465\linewidth]{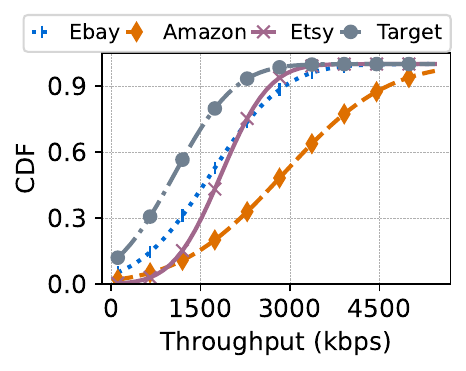}
    \label{fig:intro_a}
  }
  \subfigure[Voice Calling]{
    \includegraphics[width=0.465\linewidth]{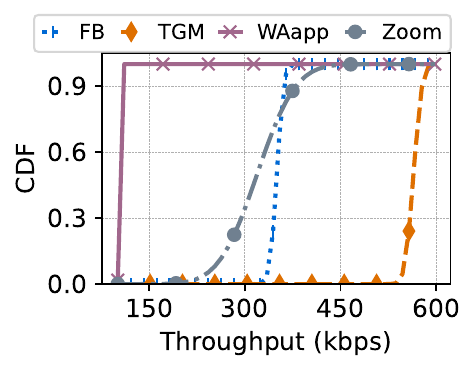}
    \label{fig:intro_b}
  }
   \caption{Comparison of 5G RRB throughput CDFs for major shopping websites \ref{fig:intro_a}; and VoIP calling apps \ref{fig:intro_b}.}
   \label{fig:intro-cdf}
\end{figure}

Our experimental results show that we can not only infer the applications from different QoS types (e.g., voice vs. web browsing traffic) but also differentiate among different applications within the same category (e.g., Target vs. Amazon). Figure~\ref{fig:intro-cdf} shows the CDF of shopping websites and voice call applications' throughput. Because various applications within one category render differently, the network schedules the RRBs according to their needs, which becomes the side channel to infer the user's activities.

In this paper, we aim to tackle the following overarching research question: ``can we infer user activity and used applications based on 5G RRB data?''. To answer this question, we collect and analyze (fingerprint) 5G network traffic and web resource data from various applications in the wild. We build a dataset consisting of the collected data, specifically, 1217 mobile traffic traces collected over six months, totaling 43GB. Our analysis of the collected data demonstrates that 5G RRB traffic can be effectively exploited to identify the applications that users run on their mobile devices. 

The contributions of our work are as follows:

\begin {itemize}

\item We collect 5G network traffic data from four different application types in the wild, i.e., online shopping, voice/video conferencing, video streaming, and over-the-top (OTT) platforms. For each application type, we consider various applications and scenarios. We thus build a comprehensive dataset, which we use for our analysis in this paper \footnote{The dataset and scripts used in this study are available at \url{https://github.com/ruman23/EncryptedTrafficAnalysis}}.

\item We analyze the collected data and fingerprint various applications. This demonstrates that 5G RRB data can be effectively exploited by adversaries to identify not only the type of user applications running on their mobile devices but also specific applications.

\end {itemize}

The rest of the paper is outlined as follows. Section \textbf{\ref{sec:radioprotocol}} provides an in-depth study of the cellular network control plane, emphasizing the RRB allocation. The subsequent Section \textbf{\ref{sec:implementation}} demonstrates our experimental setup, the data collection and processing methodology. In Section \textbf{\ref{sec:result}}, we discuss our methodologies for activity detection and key results. In Section \textbf{\ref{sec:related}}, we review recent related works on user activity identification in cellular networks and fingerprinting applications on the Internet. Finally, Section \textbf{\ref{sec:conclusion}} is the conclusion that sheds light on future research direction.

\section{Fingerprinting User Traffic 
through cellular radio Protocol}
\label{sec:radioprotocol}
The 5G physical layer is responsible for sending/receiving user data over the 5G radio channel. Whenever the user device needs to transmit data, it sends a UL scheduling request message to the network requesting for dedicated radio resources (i.e., radio resource blocks). The network replies to the device with the uplink grant, after which the device starts transmitting data over the Physical Uplink Shared Channel (PUSCH). Similarly, the device receives the DL data through DL scheduling request messages (DL grants).

Although the scheduling request message is a physical layer message, it is controlled by the MAC layer process. The MAC layer calculates the scheduling grant size from the amount of data in the MAC buffer so that the network provides enough radio resource blocks for data transmission. Further, it also implements data scheduling queues against each QoS class (i.e., resource type, priority level, packet delay budget, packet error rate, and maximum data burst) for which the network assigns the radio resource blocks. 5G device scheduler determines the radio resource blocks allocation strategy (i.e., the number and size of resource blocks) for every QoS-specific data queue whose length varies from one application to the other. There are 26 QoS Class Identifiers (QCI) indicating whether the traffic has a Guaranteed Bit Rate (GBR) or not (non-GBR) and what the class’s relative priority within those two categories is.

\begin{figure}[t]
  \centering
  \includegraphics[width=0.8\linewidth]{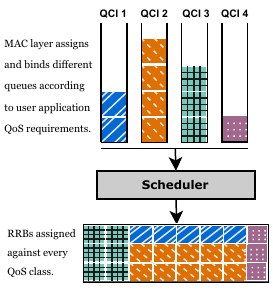}
  \caption{MAC layer assigns and binds different queues according to user application QoS requirements. The scheduler assigns RRBs against every QoS class.}
  \label{fig:resource-allocation}
  \vspace{-0.3cm}
\end{figure}

In other words, we can say that each QCI is associated with a class of traffic (often corresponding to some type of application), where a given user might be sending and receiving traffic that belongs to multiple classes at any given time. 
Figure \ref{fig:resource-allocation} shows an example where the device is running four different classes of services (e.g., voice call, web browsing, streaming service, and real-time gaming); the MAC layer allocates 4 different queues for each application type representing the particular QoS class. The scheduler takes the number, size, and priority of different queues into account when requesting radio resources from the network. The network allocates radio resource blocks, accordingly, by using which the device sends its data over the air. In this paper, we demonstrate that the attacker can fingerprint different types of applications by measuring the radio resource block size over time.

\textit{How to acquire scheduling information for device fingerprinting?}
The scheduling information is carried in DCI (Downlink Control Information) through PDCCH (Physical Downlink Control Channel). The DCI value can be either uplink (type 0) or downlink (type 1) grant for data transmission. It carries a bitmap indicating the resource block groups(RBGs) that are allocated to the scheduled UE. The DCI is transmitted without encryption over the air; the attacker can eavesdrop on the PDCCH and find the DCI scheduled for the victim. The attacker can simply consult the bitmap in DCI to locate the radio resource blocks carrying the user data.

\textit{How to identify the victim device?}
The device receives C-RNTI identity from the network as soon as it registers with the network. It is a unique ID that identifies different devices under the same cell. For example, when the network sends data packets, the device checks if the pre-coded C-RNTI matches the assigned value to get the messages intended for itself. Because the C-RNTI is sent in plain-text \cite{dabrowski2016messenger, kohls2019lost, mjolsnes2017easy, shaik2015practical}, the attacker can identify who is who for radio communication.

\noindent\textbf{Threat model:}
We consider a user (victim) accessing the Internet through mobile data. The user can access multiple online services -- apps and websites -- on their mobile device, such as those for shopping, streaming, and calling. However, we assume that they use only \textbf{one} service at any given time. This is a realistic assumption since it is impractical for a user to, for example, watch YouTube or Netflix while simultaneously making a call on the same device.

\begin{figure}[t]
  \centering
  \includegraphics[width=1\linewidth]{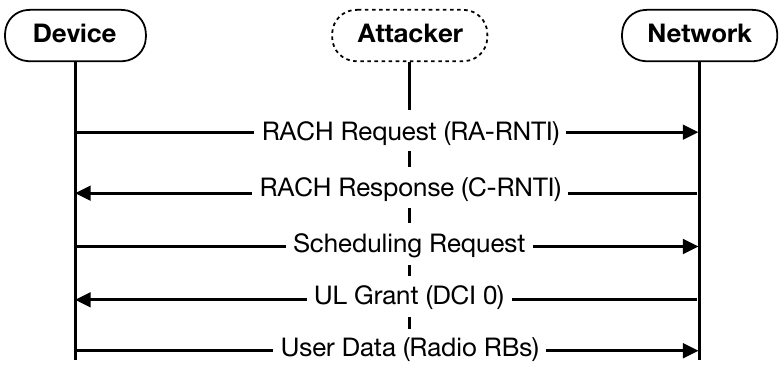}
  \caption{The device receives the C-RNTI in plaintext within the Random Access Response message from the network. The device requests the UL radio resources from the network according to its buffer size. The scheduled radio resource block information is encoded in the UL Grant message. The device sends encrypted data using the Radio Resource Blocks (RBs). The attacker knows the size and the length of RBs to plot the throughput over time and can use it as a side channel to infer the user activities.}
  \label{fig:threat-model}
\end{figure}

We consider a passive attacker who acts as a Man-in-the-Middle (MitM) eavesdropper. The attacker can passively sniff radio layer information within the victim’s cell and remain unnoticed. 
The attacker can receive and decode signals sent out by the device and the network. To do so, the attacker captures only the information that is exchanged in plain text (i.e., the communication between the device and the network, which is not encrypted). Figure~\ref{fig:threat-model} shows how the attacker can sniff plain text communication between the device and the network.

\section{Implementation}
\label{sec:implementation}
This section presents our implementation and experimental setup, followed by a discussion on the traces and applications we use for our experimental evaluation.

\noindent \textbf{Experimental setup:}
Our experiment uses a OnePlus 5G mobile phone as a victim device that only runs several GBR and non-GBR applications over a cellular connection. We collect cellular network traces through QXDM \cite{qxdm} and QCAT tools. To automate interactions with websites or applications on the OnePlus phone, we deploy Selendroid~\cite{selendroid}, the Android equivalent of Selenium. We connect the OnePlus phone to the Computer for trace collection and run QXDM and QCAT tools to collect and process Radio Resource Block traces. Mobile devices may support background applications or multitasking, but we collect the traces by ensuring the user uses a single application. 

\noindent \textbf{Applications and dataset:} We systematically analyze the 5G control plane and side channel by collecting traces from four major application categories: online shopping, social messenger, over-the-top (OTT) media services, and video streaming. We choose a set of popular applications (or activities) in each category, ensuring at least four in a category. We summarize the application categories and the specific applications in Table~\ref{tab:platforms}.

\begin{table}[htbp]
\caption{Application Categories and Specific Applications or Activities Used to Collect Mobile Traffic Traces}
\begin{center}
\begin{tabular}{|>{\centering}m{3cm}|>{\centering\arraybackslash}m{4.85cm}|}
\hline
\textbf{Category} & \textbf{Applications} \\
\hline
Online Shopping & Amazon, eBay, Etsy, Target \\
\hline
Voice/Video Conferencing & Facebook Messenger, Telegram, WhatsApp, Zoom \\
\hline
Video Streaming & YouTube (Live and Non-Live in various qualities) \\
\hline
OTT Services & Apple TV+, Amazon Prime Video, Netflix \\
\hline
\end{tabular}
\label{tab:platforms}
\end{center}
\end{table}

We collect the traces from Amazon, eBay, Etsy, and Target for online shopping. We conduct voice and video calls between two victim mobile phone users for social messenger applications such as Facebook Messenger, WhatsApp, Telegram, and Zoom. On the other hand, while collecting the traces for video streaming, mobile phones stream videos and movies of different video quality on YouTube. We stream live and non-live videos in various resolutions (SD, HD, and Full HD). Finally, we collect traces from OTT platforms while streaming video on Apple TV, Netflix, and Amazon Prime. 

Through this process, we create a dataset consisting of 1217 mobile traffic traces collected over a period of six months, confirming at least 20 or more iterations for each application, website, or activity. Instead of collecting the traces for an application at a specific time and date, we collect the traces randomly for an application to represent a wide range of user behaviors and network conditions. The overall size of our dataset is approximately 43GB.

\noindent \textbf{Data processing:}
We collect the Radio Resource Block, Wireshark Transport layers, and Web Resources traces using QXDM, Wireshark, and Selenium WebDriver's \texttt{get\_log(`performance')} method \cite{r1}, respectively. After collecting the traces, we separated the UL and DL throughput from the Radio Resource Block and Wireshark traces. We also extract the received resource types and sizes from the performance logs of Web Resources. As we collect the traces over multiple iterations, we compute their average. Finally, we normalize the Radio Resource Block and Wireshark trace on a rolling basis by applying a window size of $20\%$ of the period of the traces for further analysis. 

We remove zero values and outliers from the UL and DL throughput datasets to reduce the noise and extract features that we can use for analysis through Machine Learning (ML). First, we remove zero values and apply the Interquartile Range (IQR), defined as the difference between the 75th percentile ($Q_3$) and the 25th percentile ($Q_1$) of the data, to mitigate the effect of outliers. We apply Equation~\ref{eq:noise_adjustment} to remove the dataset's outliers. Outliers are data points lying more than 2.0 times the IQR above $Q_3$. To mitigate the impact of extreme outliers, we cap values exceeding the upper bound to a predetermined cap value set at the 95th percentile of the data.  

\begin{equation}
\text{Adjusted Value} = \begin{cases} 
\text{Cap Value}, & \text{if } x > Q_3 + 2.0 \times \text{IQR} \\
x, & \text{otherwise}
\end{cases}
\label{eq:noise_adjustment}
\end{equation}

We make use of machine learning (ML) models with the features of mean, Standard Deviation (STD), and slope values of time series data \cite{r5, faouzi2022time}. Specifically, we used the Random Forest Classifier~\cite{rf} and Extra Trees Classifier~\cite{xtree} models to the noise-free dataset on the extracted mean, STD, and slope values. We further extract the $Q_1$ and $Q_3$ values for both UL and DL. We split the extracted dataset into two parts, where 70\% of the data is used for training and 30\% is used to evaluate the performance of the models.
\section{Key Results}
\label{key_result}
Our experiment categorizes 22 applications and activities of different platforms using UL and DL Radio Resource Block (RRB) 
throughput. Besides classifying the applications and activities based on the RRB throughputs, we also characterize the shopping websites for Wireshark and received Web resource traces.

\label{sec:result}

\subsection{5G Control Plane Accurately Represents Application and Network Traffic}
\label{subsec:5g-control}

\begin{figure}[h!t]
  \centering
  \includegraphics[width=0.98\linewidth]{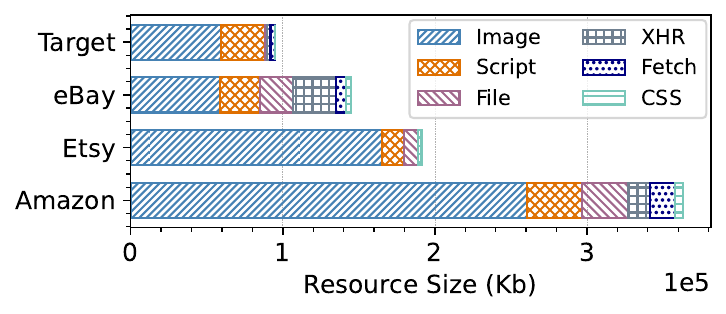}
  \caption{Distribution of the top six resources downloaded across four major shopping websites. Each website uses a unique combination of resources of varying sizes.}
  \label{fig:resouces-size}
\end{figure}

Our analysis shows that each shopping website creates a unique footprint based on the number of web resources it downloads, their types, and their size for accomplishing similar types of activity. For example, Amazon downloads about 47MB of resources, significantly more than Etsy, eBay, and Target (specifically about 49\%, 60\%, and 74\% more, respectively). In Figure~\ref{fig:resouces-size}, we show the detailed breakdown of the top six received resource types and their sizes for each website. The counts and measures of resources and their received time always vary because of websites and servers' design patterns and development approaches. Resource utilization always contributes to creating a unique identity for each website. Besides that, our analysis shows that RRB and Wireshark throughput are positively correlated to the received resources. Resource utilization is distinctive to a website, generating unique RRB and Wireshark throughput.

\begin{figure}[h!]
	\vspace{-0.2cm}
	\begin{tcolorbox}[boxsep=1pt,left=5pt,right=5pt,top=5pt,bottom=5pt]
		\textbf{Insight 1:}
  Each website generates unique footprints based on the numbers, types, and sizes of downloaded resources.
	\end{tcolorbox}
	\vspace{-0.3cm}
	\label{fig:issue1}
\end{figure}

\begin{figure}[ht]
  \centering
  \includegraphics[width=\linewidth]{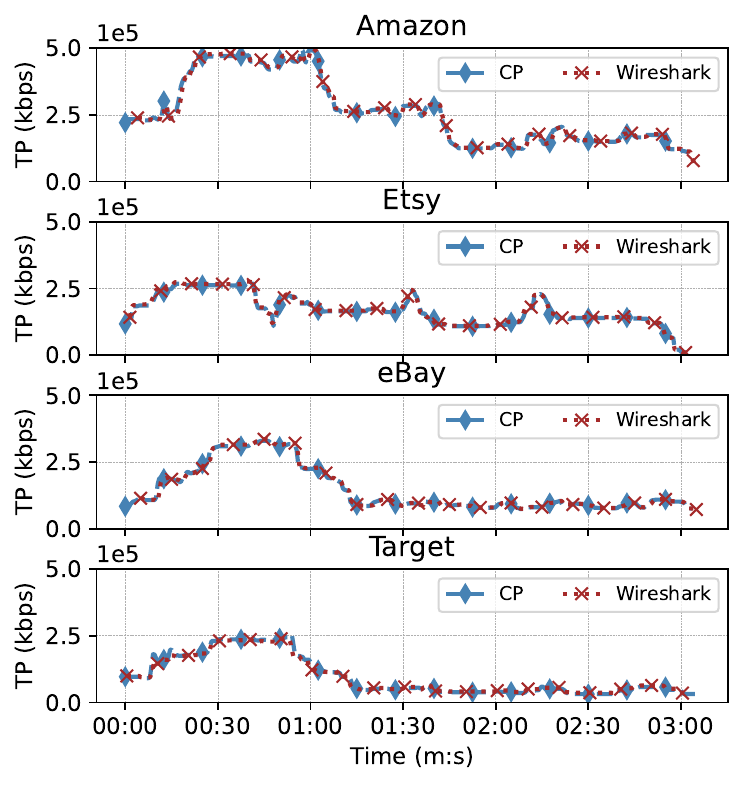}
  \caption{Normalized RRB Control plane (CP) and Wireshark downlink (DL) for major shopping websites have identical throughput (TP) patterns. The observation holds true across multiple time instances. For this plot, Wireshark throughput is scaled by a factor of 0.2.}
  \label{fig:comparison_of_pdcp_and_wireshark}
\end{figure}

We compare the RRB DL with Wireshark to determine whether both exhibit similar throughput patterns. Our analysis shows that both throughputs are identical twins, and their pattern exactly matches each other, which we show in Figure~\ref{fig:comparison_of_pdcp_and_wireshark}. We can use the RRB traces instead of Wireshark traces to classify and identify the applications and user activity as they generate a similar pattern.

\begin{figure}[h!]
	\begin{tcolorbox}[boxsep=1pt,left=5pt,right=5pt,top=5pt,bottom=5pt]
		\textbf{Insight 2:} 
  RRB and Wireshark throughputs exhibit the same patterns, allowing RRB traces to be used in lieu of Wireshark for application and activity identification.
	\end{tcolorbox}
	\vspace{-0.5cm}
	\label{fig:issue2}
\end{figure}

\begin{figure}[htb]
  \centering
  \includegraphics[width=1\linewidth]{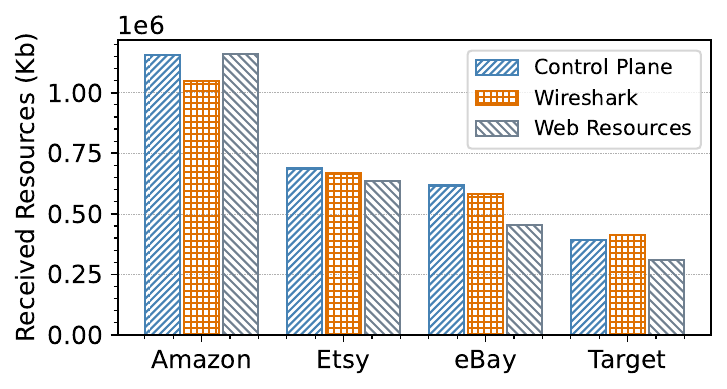}
  \caption{Comparison of the total resources of Control Plane, Wireshark, and Web Resources for shopping websites. The Wireshark and Web resources are scaled by factors of 0.2 and 3.25, respectively.}
  \label{fig:pdcp_wireshark_resource}
\end{figure}

We also compare the total received Resources with RRB and Wireshark DL throughput to determine whether they are related or not, which we demonstrate in Figure~\ref{fig:pdcp_wireshark_resource}. According to the graph, the total received traffic of three throughputs is correlated for a website, but their values vary on different websites. So, received Resources lead to increases and decreases in the RRB and Wireshark throughput, and we have already discussed that received resources are responsible for unique footprint. As a result, RRB and Wireshark throughput will always be unique for any website, application, or activity. 

\begin{figure}[h!]
	\vspace{-0.2cm}
	\begin{tcolorbox}[boxsep=1pt,left=5pt,right=5pt,top=5pt,bottom=5pt]
	\textbf{Insight 3:} 
   A correlation exists among total received resources, RRB, and Wireshark throughput, indicating the unique characteristics of applications and activities.
	\end{tcolorbox}
	\vspace{-0.3cm}
	\label{fig:issue3}
\end{figure}

\begin{figure*}
    \centering
    \subfigure{
        \includegraphics[width=0.31\textwidth]{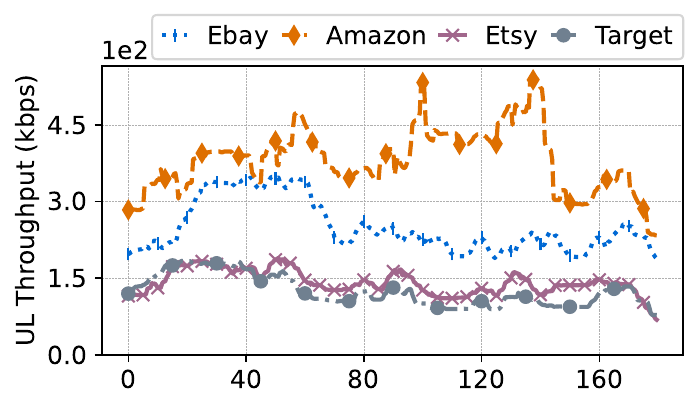}
    }
    \subfigure{
        \includegraphics[width=0.31\textwidth]{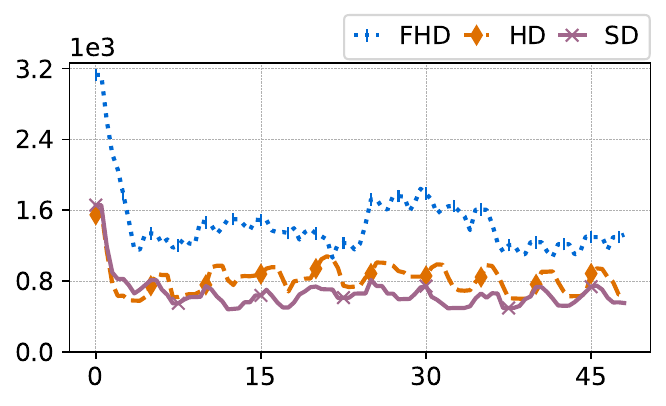}
    }
    \subfigure{
        \includegraphics[width=0.31\textwidth]{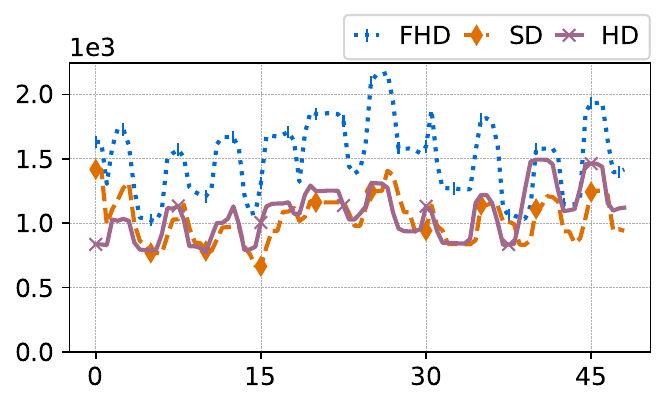}
    }\\
    \vspace{-0.25cm}
    \subfigure[Shopping Websites]{
        \setcounter{subfigure}{1}
        \includegraphics[width=0.32\textwidth]{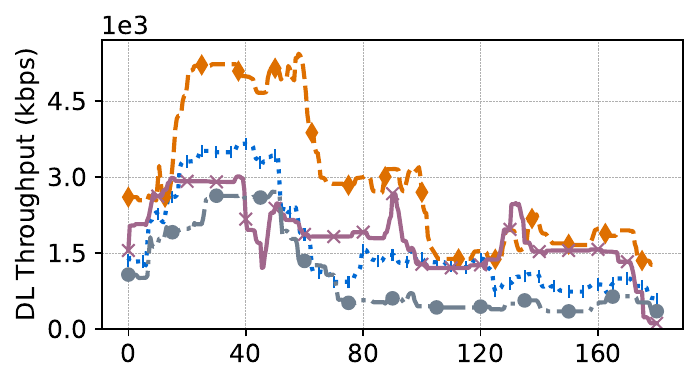}
        \label{fig:throughput_a}
    }
    \subfigure[YouTube Live]{
        \includegraphics[width=0.31\textwidth]{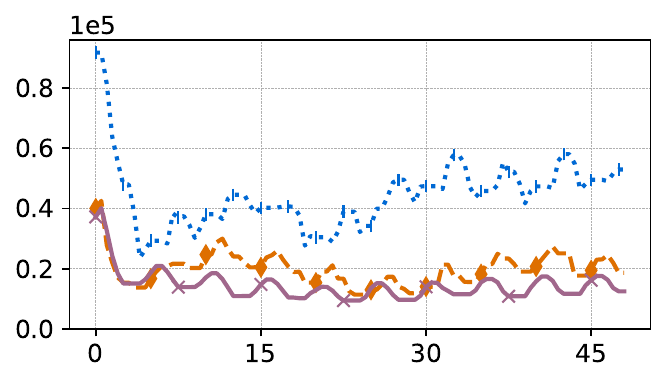}
        \label{fig:throughput_b}
    }
    \subfigure[YouTube Non-Live]{
        \includegraphics[width=0.31\textwidth]{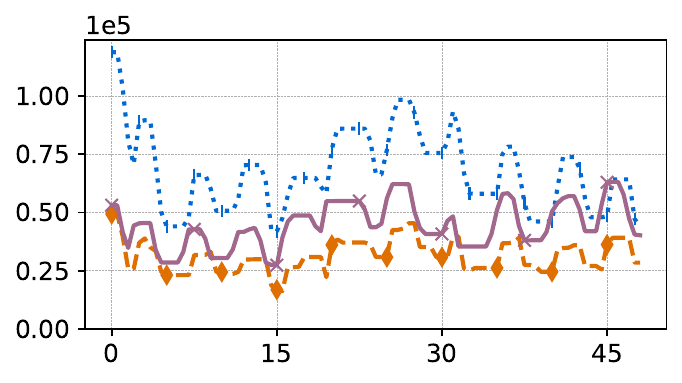}
        \label{fig:throughput_c}
    }\\

    \subfigure{
        \includegraphics[width=0.32\textwidth]{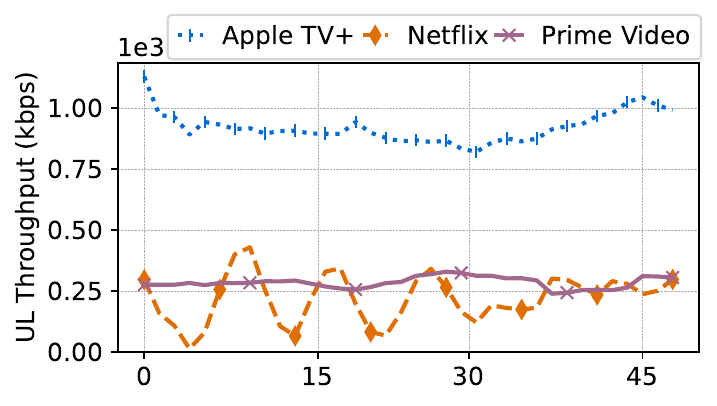}
    }
    \subfigure{
        \includegraphics[width=0.31\textwidth]{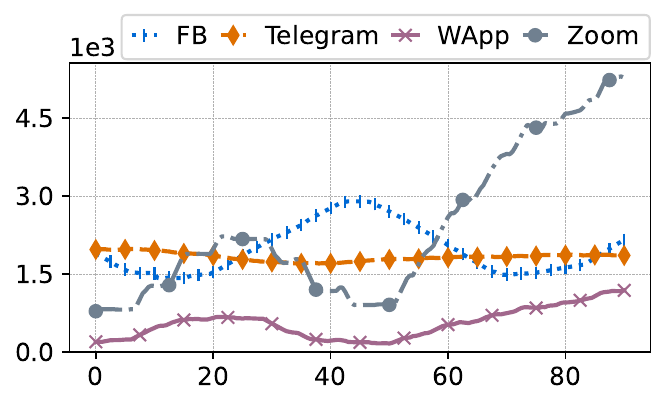}
    }
    \subfigure{
        \includegraphics[width=0.31\textwidth]{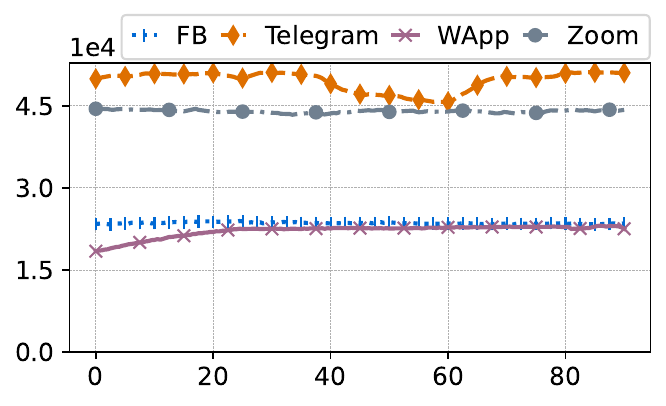}
    }\\
    \vspace{-0.25cm}
    \subfigure[OTT Platforms]{
        \setcounter{subfigure}{4}
        \includegraphics[width=0.32\textwidth]{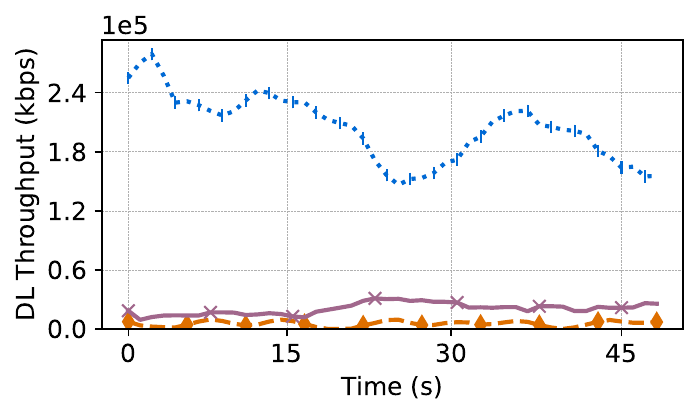}
        \label{fig:throughput_d}
    }
    \subfigure[Voice Call]{
        \includegraphics[width=0.31\textwidth]{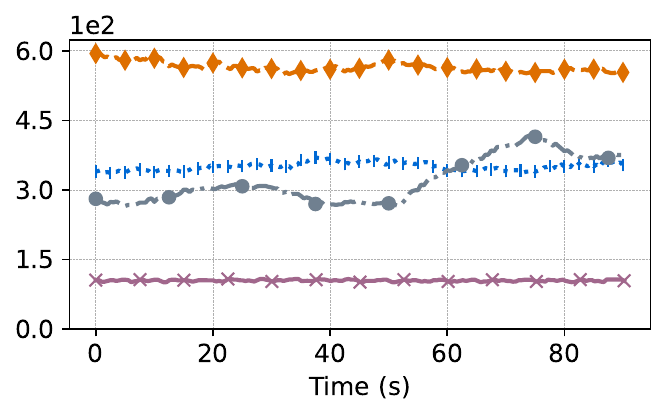}
        \label{fig:throughput_e}
    }
    \subfigure[Video Call]{
        \includegraphics[width=0.31\textwidth]{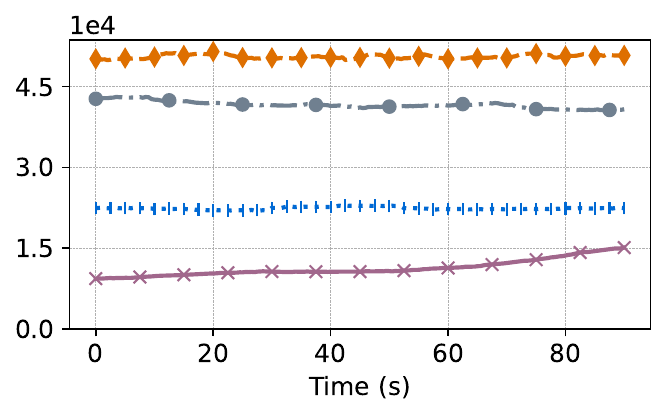}
        \label{fig:throughput_f}
    }

    \caption{Normalized RRB throughput for different applications and their activities where the top graph represents UL and the bottom represents DL throughput. In (b) and (c), SD, HD, and FHD indicate standard, high, and full HD definition qualities, respectively. The throughput is normalized on a rolling basis where window size equals 20\% of the time period.}
    \label{fig:throughput}
\end{figure*}

\subsection{Fingerprinting Applications in the Wild}

We analyze the collected traffic traces to identify application types, specific applications, and activities from the encrypted RRB UL and DL control plane traffic. In Figure~\ref{fig:throughput}, we present both RRB UL and DL throughput for various applications and scenarios.
Figure~\ref{fig:throughput_a} presents throughput for the four shopping websites. Similarly, we contrast the throughput of YouTube for streaming both live (Figure~\ref{fig:throughput_b}) and non-live (Figure~\ref{fig:throughput_c}) videos in standard (SD), high (HD) and full (FHD) definition qualities.
Figure~\ref{fig:throughput_d} presents the throughput for OTT platforms. Finally, in Figures\ref{fig:throughput_e} and Figure~\ref{fig:throughput_f}, we present various scenarios when two users, a caller, and a callee, communicate through voice/video conferencing applications, specifically, WhatsApp (WApp), Zoom, Facebook Messenger (FB) and Telegram.

The patterns and shapes of throughput over time differ from one application type to another. Typically, YouTube generates repeated sinusoidal patterns; OTT applications generate convex patterns; conferencing applications generate linear patterns; and shopping websites generate patterns distinct from other platforms as well. Our results indicate that DL throughput is typically higher than UL as users consume more content (like web pages, videos, and images) than they upload. As such, UL throughput is lower because user-generated data sent to the servers, such as clicks, text inputs, or occasional uploads, is smaller in terms of size. In contrast, for voice and video calls, UL throughput is higher than the DL throughput because we collect the traces on the end of the user actively talking on the phone.

\noindent \textbf{Online shopping:} For shopping websites, user activity and website design influence the UL and DL throughput variability, as outlined in Section~\ref{subsec:5g-control}. High initial DL throughput is attributed to downloading resource-intensive web elements like images and scripts. Over time, this demand decreases as resources are cached and reused, reducing the need for repeated downloads. The UL throughput varies based on user interactions, often sending activity-tracking data back to web servers. Our results (Figure~\ref{fig:throughput_a}) demonstrate that Amazon has the highest DL and UL throughput among shopping applications since Amazon results in downloading and uploading the most significant volume of web resources as shown in Section~\ref{subsec:5g-control}. On the other hand, Target has the lowest DL and UL since Target results in downloading and uploading the smallest volume of web resources, as shown again in Section~\ref{subsec:5g-control}. Etsy and Target appear to generate similar UL throughput graphs. However, the STD of UL throughput for Target is more than 55\% higher than Etsy, and  Etsy results in about 16\% higher median UL throughput than Target.
%

\noindent \textbf{Video streaming:} For video streaming, UL throughput is lower than DL since a substantial volume of video content is downloaded, while a minimal amount of data is uploaded to maintain the connection and the streaming quality. Throughput variability, with its peaks and troughs, derives from adaptive streaming protocols that adjust video quality to control frame buffering and manage network congestion. Live streaming typically shows a more consistent throughput than non-live streaming, as it maintains a real-time data flow, while non-live content, already stored on servers, can be buffered to accommodate network changes.
Our results (Figures~\ref{fig:throughput_b} and~\ref{fig:throughput_c}) indicate that we can differentiate between live and non-live video streaming and streaming qualities. Specifically, non-live video streaming results in comparatively higher DL throughput than live, while DL throughput increases with the video quality.

\noindent \textbf{OTT platforms:} Our analysis indicates that OTT platforms generate lower UL throughput than DL throughput due to the fact that user devices primarily download video frames while they upload rather small amounts of data to acknowledge received frames and maintain the connection. Our results (Figure~\ref{fig:throughput_d}) demonstrate that Apple TV+ results in the highest DL throughput since the default video quality used by this application is higher (up to 4K HD) than the default video quality used by Amazon Prime Video and Netflix. We can differentiate among the different OTT media platforms based on both the UL and DL throughput.


\noindent \textbf{Voice/video conferencing:} The UL and DL throughputs in voice conferencing applications vary depending on whether the caller or the callee is talking. As the caller talks intermittently during a voice call, there is variability in UL throughput. However, this variability contrasts with the more consistent DL throughput because the callee is not talking, but applications constantly transmit a minimum amount of data to maintain the connection regardless of a user's speech activity. Both UL and DL are more consistent in video calls than in voice calls due to the continuous nature of video streaming, where data is transmitted at a constant frame rate. For both voice and calls, throughput can vary from one application to another depending on the application development process and data compression process. 
Our results (Figures~\ref{fig:throughput_e}-\ref{fig:throughput_f}) indicate that we can distinguish among voice/video conferencing applications. For example, in the case of video calls, Telegram and Zoom consistently generate the highest and second-highest throughputs over time, while Facebook and WhatsApp generate the lowest throughputs. Regarding voice calls, Zoom generates the highest UL throughput, followed by Facebook, Telegram, and WhatsApp. 

Finally, we analyze the traffic traces in a time-independent manner by plotting the Cumulative Distribution Function (CDF) of the encrypted RRB UL and DL control plane traffic. Our results (presented in Figure~\ref{fig:cdf} in the appendix) demonstrate the same behavior as our results presented in Figure~\ref{fig:throughput}.

\begin{figure}[h!]
\vspace{-0.2cm}
\begin{tcolorbox}[boxsep=1pt,left=5pt,right=5pt,top=5pt,bottom=5pt]
    \textbf{Insight 4:} 
Throughput distribution of encrypted RRB UL and DL control plane traffic can be exploited to differentiate among different applications.
\end{tcolorbox}
\vspace{-0.3cm}
\label{fig:issue4}
\end{figure}

\subsection{Evaluating the Effectiveness of RRB Throughput for Application Categorization Using ML}

In addition to fingerprinting applications using throughput and CDF graphs, we use the Random Forest Classifiers and Extra Trees Classifiers to classify the applications and activities. We use the same hyperparameters to ensure a uniform experimental setup and to compare the performance of the classifiers directly. Specifically, we set 100 for the \texttt{n\_estimators}, and this parameter is essential as it determines the number of trees in the forest, influencing the balance between computational efficiency and model accuracy. Furthermore, we set 42 for the \texttt{random\_state}, and this parameter serves as a seed for the random number generator. It ensures reproducibility by adding determinism to the stochastic processes inherent in these ML methods. The performance of both models is shown in Table~\ref{tab:performance_comparison}. Our analysis shows that we can effectively classify applications in the wild by utilizing the extracted features (mean, STD, slope, $Q_1$ and $Q_3$ of UL and DL) of the RRB throughput. The Random Forest Classifier can uniquely identify an activity or an application with 94\% accuracy, whereas the Extra Trees Classifier achieves an accuracy of 90\%. These results demonstrate that we can use the RRB throughput as a reliable data source for identifying and categorizing activities and applications.

\begin{table}[h]
\caption{Performance metrics of ML models (Random Forest Classifier and Extra Trees Classifier).} 
\label{tab:performance_comparison}
\begin{tabular}{|>{\centering\arraybackslash}p{3cm}|>{\centering\arraybackslash}p{2.2cm}|>{\centering\arraybackslash}p{2.2cm}|}
\hline
\textbf{Metric} & \textbf{Random Forest}      & \textbf{Extra Trees}           \\ \hline
Accuracy            & 94\% & 90\% \\ \hline
Macro Average Precision & 93\% & 91\% \\ \hline
Macro Average Recall    & 94\% & 93\% \\ \hline
Macro Average F1-Score  & 93\% & 91\% \\ \hline
Lowest F1-Score & 86\% & 67\% \\ \hline
\end{tabular}
\end{table}

Besides analyzing the models' performance based on their accuracy, we also calculate the precision, recall, and F1-score independently for each application/activity category, then average these scores across all categories without considering the class imbalance, which is known as the macro average method \cite{liu2022new}. This approach provides a broad perspective on the classifiers' performance, ensuring that all classes contribute equally to the final average regardless of iteration size.

Though the Random Forest Classifier's lowest F1 score for identifying a category (YouTube live streaming in SD quality in particular) is 86\%, the macro average precision, recall, and F1 score are 93\%, 94\%, and 93\% respectively, reflecting a high level of consistency and balance in performance across different categories. Similarly, the lowest F1 score for identifying a category (YouTube live streaming in SD quality) is 67\% for the Extra Trees Classifier model, but the macro average precision and recall are 91\% and 93\%, respectively, with a macro average F1-score of 91\%. These metrics indicate that, despite slight variations, both models leverage RRB throughput effectively to classify applications, with each category being accurately recognized based on its unique throughput characteristics.

 \begin{figure}[h!]
\vspace{-0.2cm}
\begin{tcolorbox}[boxsep=1pt,left=5pt,right=5pt,top=5pt,bottom=5pt]
    \textbf{Insight 5:} 
ML models can effectively use features, such as the mean, STD, slope, $Q_1$, and $Q_3$, from the throughput distribution of encrypted RRB UL and DL control plane traffic to identify applications and activities with high performance.
\end{tcolorbox}
\vspace{-0.3cm}
\label{fig:insight5}
\end{figure}
\section{Related Work}
\label{sec:related}

\noindent \textbf{Identifying user activities through cellular networks:} Closest to our work are \cite{baek2023targeted, kohls2019lost, ludant20235g, rupprecht2019breaking}. In \cite{rupprecht2019breaking}, the attacker utilizes LTE layer 3 for fingerprinting websites, whereas \cite{kohls2019lost} uses metadata of layer two for determining users' accessed website behavior. \cite{rupprecht2019breaking,kohls2019lost} fail to fingerprint websites that belong to the same class (e.g., e-commerce websites). Further, their approach is neither real-time nor differentiates the applications from different QoS classes (e.g., GBR and non-GBR). In contrast, this paper uses 5G physical layer protocol operations as a side channel to identify both GBR and non-GBR applications in real-time. \cite{ludant20235g} collects Radio Network Temporary Identifiers (RNTIs) for fingerprinting messaging applications such as Signal and Telegram. However, their process depends on the victim's virtual identity, such as social media handles or email addresses, and requires a particular app to interact with the known virtual identity. Our approach, in contrast, can fingerprint several applications, including online shopping, video streaming, and calling, without prior user information.
\cite{baek2023targeted}  studies side-channel attacks on mobile apps (streaming, messaging, and VoIP) requiring application-specific features as a pre-condition. In contrast, we do not require knowledge of the application; rather, our approach looks at 5G signaling to detect the application behavior.  
\cite {lichtman2013vulnerability, jover2013security} discuss DoS attacks on the cellular physical layer, whereas this paper uses the physical layer as a side channel for inferring user activities. 

\noindent \textbf{Fingerprinting applications on the Internet:} \cite{cheng1998traffic} leads the exploration of traffic analysis vulnerabilities in SSL-encrypted web browsing, showcasing the potential of traffic fingerprinting techniques to identify individual web pages. Several other attempts \cite{ r19, r21, r17, r18} have focused on leaking user information from the encrypted traffic.
 
\cite{r12} circumvents traffic obfuscation using IP-to-IP communication graph and IP host information to profile different application groups from network traffic. This approach requires prior IP host data and other traffic classifiers for seed information. 
Other works \cite{r13, r14, r15} have researched on payload inspection paired with clustering algorithms to classify internet traffic.
The rise in the popularity of machine and deep learning algorithms has given a new direction to online application fingerprinting. Notably, \cite{r19, r17, r4, r6, zander2005automated, r20} use statistical models and deep learning algorithms to profile 
users~\cite{anwar2023detecting} and their activity
traffic to identify applications using features like packet size, direction, time intervals, and bytes distribution. However, these methods are not generalized and mainly work for limited types of applications, i.e., instant messaging, mail, and file transfer. 

Unlike all these works, we exploit the 5G layer one and two interaction to fingerprint real-time user traffic. 
Our approach utilizes the physical layer's RRB allocation process in determining different application types (e.g., shopping website vs. streaming), classifying the exact application (e.g., Netflix vs. Disney+), identifying the activity (e.g., audio vs. video call), and identifying the streaming types (live vs. non-live). 
\section{Conclusion and Future Work}
\label{sec:conclusion}

Despite the widespread use of encryption in network communications today, including mobile communications, fingerprints generated by mobile applications can be exploited by adversaries to identify user activities and the applications themselves. In this paper, we fingerprinted various mobile applications in the wild and showed that 5G control plane traffic (specific RRB data) can be effectively exploited to identify not only the type of the applications running but also the specific applications. Our work highlighted the following insights: (i) mobile applications generate unique footprints based on the number, types, and sizes of downloaded and uploaded resources; (ii) a correlation exists among the total resources, RRB, and Wireshark throughput, indicating the unique characteristics of applications and user activities; and (iii) we can analyze both the time-dependent and time-independent distributions of encrypted RRB UL and DL control plane traffic to differentiate among different applications types as well as specific applications for a particular type. 

Our study serves as a foundation for future investigations on how 5G control plane traffic can be exploited to compromise user security and privacy. Ultimately, our study aims to inform future design, implementation, and deployment decisions of 5G mobile networks and beyond, as well as motivate the realization of defense mechanisms against cellular control plane attacks.  

\section*{Acknowledgements}

This work is partially supported by the National Science Foundation through awards CNS-2345563, CNS-2104700, and CNS-2306685, as well as by the U.S. Army Engineer Research \& Development Center (ERDC) through HPC PET special project 5025.
\bibliographystyle{IEEEtran}
\bibliography{references}
\appendix

\begin{figure*}[b]
    \centering
    \subfigure{
        \includegraphics[width=0.32\textwidth]{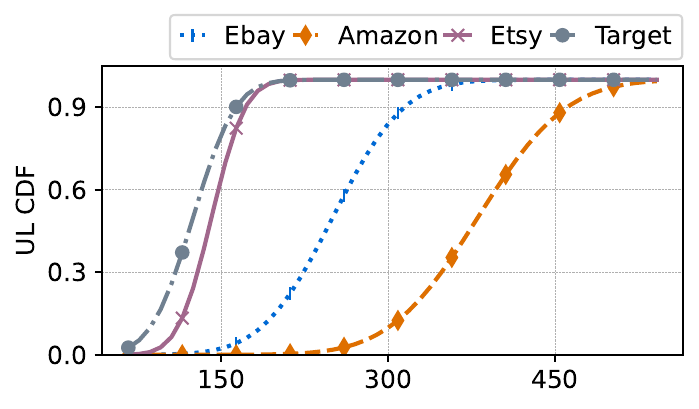}
    }
    \vspace{-0.3cm}
    \hfill
    \subfigure{
        \includegraphics[width=0.31\textwidth]{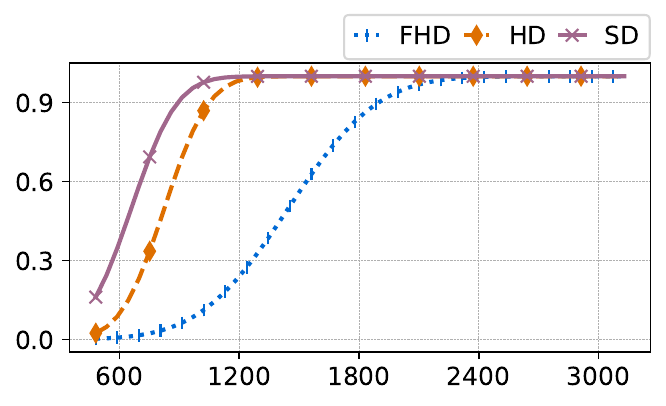}
    }
    \hfill
    \subfigure{
        \includegraphics[width=0.31\textwidth]{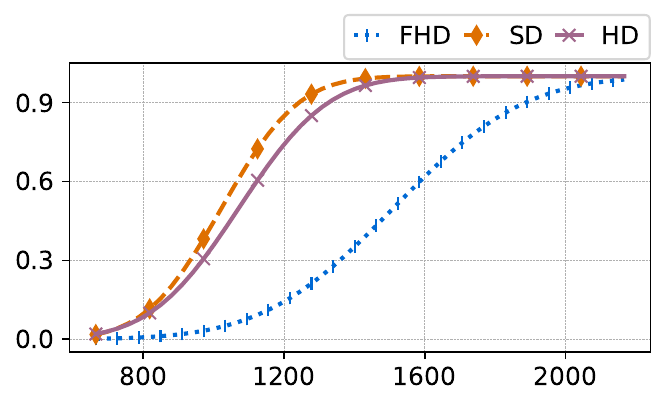}
    }
    \subfigure[Shopping Websites]{
    \setcounter{subfigure}{1}
        \includegraphics[width=0.32\textwidth]{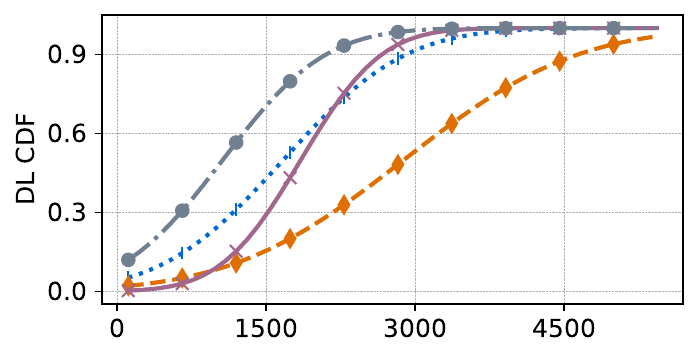}
        \label{fig:cdf_a}
    }
    \vspace{-0.1cm}
    \hfill
    \subfigure[YouTube Live]{
        \includegraphics[width=0.31\textwidth]{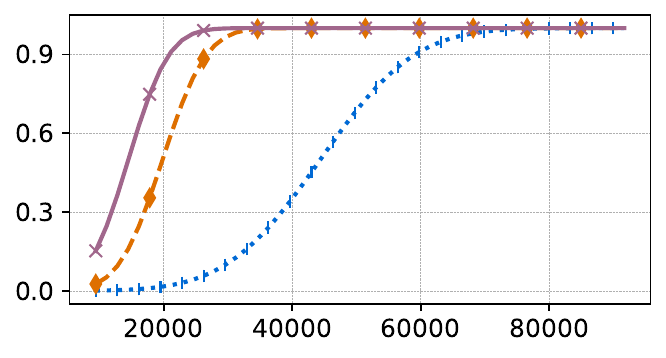}
        \label{fig:cdf_b}
    }
    \hfill
    \subfigure[YouTube Non-Live]{
        \includegraphics[width=0.31\textwidth]{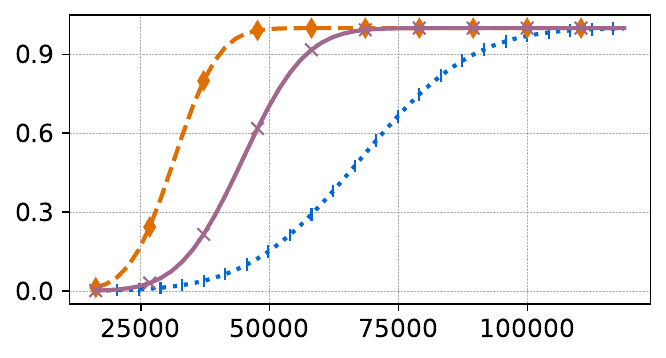}
        \label{fig:cdf_c}
    }
    \vspace{-0.3cm}
    \subfigure{
        \includegraphics[width=0.32\textwidth]{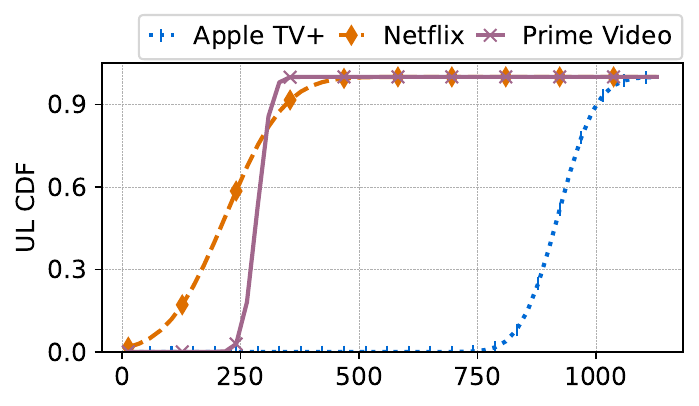}
    }
    \hfill
    \subfigure{
        \includegraphics[width=0.31\textwidth]{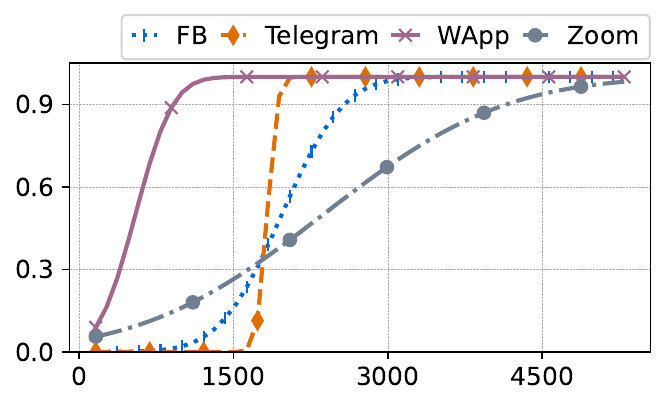}
    }
    \hfill
    \subfigure{
        \includegraphics[width=0.31\textwidth]{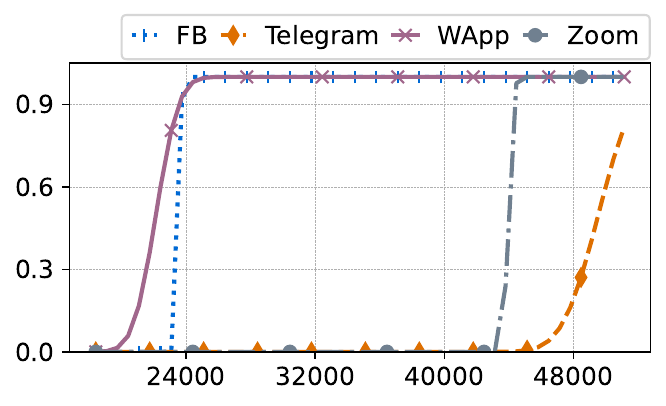}
    }

    \subfigure[OTT Platforms]{
    \setcounter{subfigure}{4}
        \includegraphics[width=0.32\textwidth]{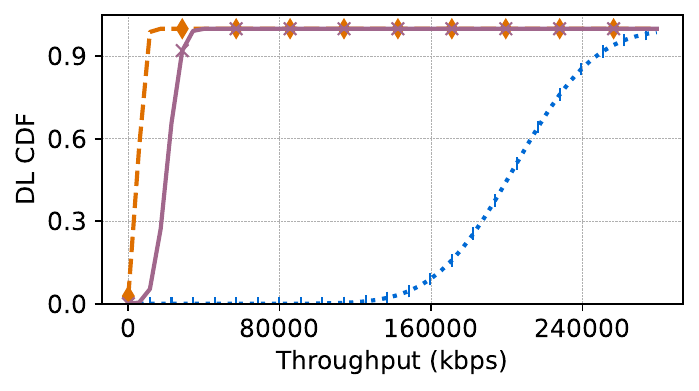}
        \label{fig:cdf_d}
    }
    \hfill
    \subfigure[Voice Call]{
        \includegraphics[width=0.31\textwidth]{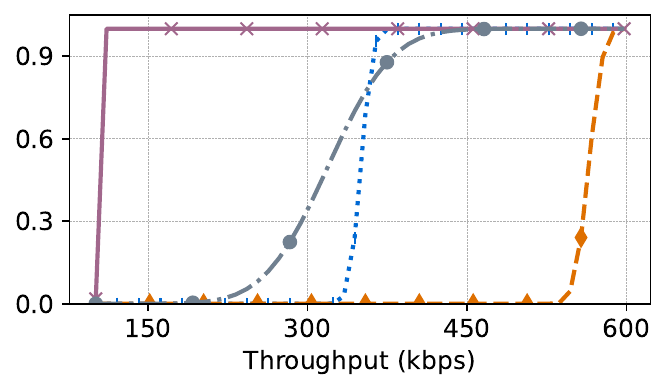}
        \label{fig:cdf_e}
    }
    \hfill
    \subfigure[Video Call]{
        \includegraphics[width=0.31\textwidth]{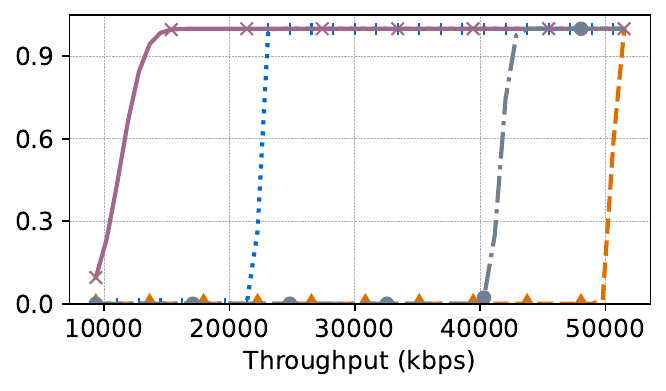}
        \label{fig:cdf_f}
    }
    \caption{CDF graphs on normalized RRB throughput of different applications and their activities where the top graph represents UL and the bottom represents DL throughput. In (b) and (c), SD, HD, and FHD indicate standard, high, and full HD definition qualities, respectively.}
    \label{fig:cdf}
\end{figure*}

The throughput of applications or websites depends on a user's activity. If there is no activity, for example, in shopping or OTT platforms, then the throughput will be close to zero. If we remove the throughput data points close to zero, the total throughput of performing an activity is impacted by the distribution of non-zero data points. 
To this end, in Figure~\ref{fig:cdf}, we present the CDF of the encrypted RRB UL and DL control plane traffic for various applications. Our results indicate that we can use this time-independent distribution of UL and DL RRB throughput to distinguish among different application types as well as specific applications for a particular type.
\balance
\end{document}